\documentclass[twocolumn,pre]{revtex4}

\usepackage[final]{graphics}
\usepackage{amssymb}
\usepackage{amsmath}
\usepackage{latexsym}

\begin{document}


\title{Charge Regulation of Interacting Weak Polyelectrolytes\footnote{
\textit{J. Phys. Chem. B} (2004), in press.}}
\author{Yoram Burak}
\email{yorambu@post.tau.ac.il}
\affiliation{School of Physics and Astronomy, \\
Raymond and Beverly Sackler Faculty of Exact Sciences \\
Tel Aviv University, Tel Aviv 69978, Israel}
\author{Roland R. Netz}
\affiliation{Sektion Physik, LMU Munich, Theresienstr. 37, 80333 
Munich, Germany}
\date{November 03, 2003}

\begin{abstract}
We introduce a generalized non-uniform mean-field formalism to 
describe the dissociation of weak rod-like polyelectrolytes (PEs).
Our approach allows for two-sublattice symmetry breaking 
which in titration curves is associated with a plateau 
for intermediate dissociation degrees.
We first test our method in the case of a single weak PE
by comparison with exact enumeration studies and show that 
 it gives quantitatively
accurate results for the dissociation degree in the full range
of pH values, and in specific performs much better than
the nearest-neighbor approximation (where exact solutions are possible).
We then study charge regulation of the coupled
system of a  weak polyacid and a weak polybase as a function 
of their mutual distance, which has some relevance for 
PE-multilayer formation and for PE complexation. An intricate interplay
of the degree of dissociation and the effective interaction
between the PEs as a function of their mutual distance
is found.
\\
\end{abstract}

\maketitle


\section{Introduction}
 
The charge of weak polyacids and polybases is determined 
by the probability of each functional group to dissociate and expose a 
charged residue. This probability depends on a chemical equilibrium 
which can be tuned by varying the pH of the solution. 
In contrast to dilute solutions of monoacids or monobases, 
in weak polyelectrolytes (PEs)
each functional group is influenced by all other groups 
along the polymer, via their mutual 
electrostatic interaction. 
As a result of the repulsion between charged groups, even strong PEs 
become weak at low salt concentrations.
Furthermore, when two or more polymers 
interact with each other, their degree of ionization is modified, 
compared to their isolated state, and depends on parameters
such as the distance between the polymers and their relative 
spatial configuration. Due to the many-body nature of this problem, 
and the long range of the electrostatic interactions, an exact solution 
for the average charge as function of pH is generally not known. 

In this paper we consider stiff PEs, where there is no 
coupling between the dissociation degree of freedom and the 
polymer conformation (for treatment of such coupling in flexible PEs see,
for example, \cite{Raphael,Borukhov,Zito}). 
We consider first a single PE 
in salt solution and discuss some of the approximations 
commonly used to characterize its charge regulation. 
It was previously shown that a uniform  mean field approach 
cannot adequately describe charge regulation when the 
coupling between charges along the polymer is strong \cite{Borkovec}. 
In these cases ionizable groups dissociate in a two-step process, 
characterized by a plateau in the charge \textit{vs.} pH curve. 
This process results from a spatially inhomogeneous charging pattern and 
is not predicted by a uniform mean field approach. 
We introduce a mean field theory
with explicit symmetry breaking between two sublattices.
Such an approximation is shown to be semi-quantitatively 
accurate and performs better than previous calculations 
where the range of interactions is restricted \cite{Borkovec},
as we demonstrate by comparison with exact enumeration over all
configurations for finite chain lengths. 

In the second part of the paper we apply our non-uniform mean-field 
scheme to the interaction of two
stiff PEs. We restrict ourselves to the simple case
of polymers aligned parallel to each other and calculate the average
charging and free energy as function of their distance. Our model
reveals some of the intricate effects that can occur in interacting
weak PEs. In a
broader context, these interactions are of interest
in the formation of PE multilayers, composed
of alternating layers of positively and negatively charged 
polymers \cite{Decher,Donath,Caruso}. 
In particular weak polyacids and polybases
can be used to form multilayers \cite{Yoo98,Rubner00}.
In this case properties
such as the layer thickness and density depend strongly on the
dissociation degree of the functional groups, and can be 
tuned sensitively by varying the pH of the solution \cite{Rubner00}.


\section{Single Polyelectrolyte}
The free energy for a weak PE, immersed in
an aqueous ionic solution, can be written as follows:
\begin{equation}
F = -{\rm ln}\sum_{\{s_i=0,1\}}{\rm exp}(-\mathcal{H})
\label{partitionf}
\end{equation}
The Hamiltonian $\mathcal{H}$ is given by:
\begin{equation}
\mathcal{H} = \mu \sum_{i} s_i + \sum_{i>j}v_{\rm DH}({\bf r}_i,{\bf r}_j) 
\label{hamiltonian}
\end{equation}
where $r_i$ is the position of the $i$th monomer and
$s_i$ can be either zero (for an uncharged monomer) or one (charged
monomer). 
The sum in Eq. (\ref{partitionf}) goes over all different configurations of
dissociated groups. Note that $F$ and $\mathcal{H}$ are given in units
of the thermal energy $k_BT$.
The chemical potential $\mu$ 
is related to the pH of the solution
\cite{Netz03}:
\begin{eqnarray}
\mu & = & -2.303 ({\rm pH}-{\rm pK_a}) - l_B \kappa 
\ \ \ \ \ \ \ \ \mbox{(Acid)} \nonumber \\
\mu & = & 2.303 ({\rm pH} - {\rm pK_b}) - l_B \kappa 
\ \ \ \ \ \ \ \ \ \ \mbox{(Base)} \label{mu}
\end{eqnarray}
where $\kappa$ is the Debye screening length,
$l_B = e^2/(\varepsilon k_B T)$ is the Bjerrum length, equal
to about 7\,\AA~in water at room temperature, $k_B T$ is the thermal 
energy, $\varepsilon_w$ is the dielectric constant of water and $e$ is
the unit charge. 
The last term in Eq. (\ref{mu}) is the self-energy of the two charges
created in the dissociation process.
We assume 
throughout this paper that the ionic solution can be described 
using the linearized Debye-H\"{u}ckel theory, so that electrostatic
interactions between charges are pairwise additive, as in
Eq.~(\ref{hamiltonian}). The exact form of $v_{\rm DH}$ 
depends on the salt concentration,
and also on the dielectric properties of the polymer backbone, as will
be discussed below. In the most simple case of dielectric continuity
between the polymer and solution, $v_{\rm DH}$ is equal to:
\begin{equation}
v_{\rm DH}({\bf r}_1,{\bf r}_2) = l_B 
\frac{{\rm e}^{-\kappa \left| {\bf r}_1-{\bf r}_2\right|}}
{\left| {\bf r}_1-{\bf r}_2\right|}
\label{vDH}
\end{equation}
The linear Debye-H\"uckel approach neglects non-linear effects that are
associated with counterion condensation and which are contained in the
non-linear Poisson-Boltzmann formalism. The main reason for resorting
to linear theory is that only at that level can the complicated problem of
spatially inhomogeneous charge distributions on the PE backbone be calculated.
One justification is that weak polyelectrolytes as studied in this paper
are typically not strongly charged, so that non-linear effects are less important
than for strong polyelectrolytes, as will be discussed in more detail in the 
concluding section.

For the following calculations,
it is convenient to use symmetric variables $\tilde{s}_i$ having
the values $-1, 1$ instead of $0, 1$:
\begin{equation}
s_i = \frac{1+\tilde{s}_i}{2} 
\end{equation}
In terms of these variables the partition function is:
\begin{eqnarray}
Z & = & \sum_{\{\tilde{s}_i=-1,1\}}{\rm exp}
\left\{
-\tilde{c} - \tilde{\mu}\sum_{i}{\tilde{s}_i} 
\right. \nonumber \\
& & \left.
-\sum_{i>j}\tilde{s}_i\tilde{s}_j\tilde{v}_{\rm DH}
\left[a(i-j)\right]\right\}
\label{znew}
\end{eqnarray}
where:
\begin{eqnarray}
\tilde{c} & = & \frac{1}{2} N \mu + \frac{1}{4} N \sum_{j>0}
                  v_{\rm DH}[a j] \nonumber \\
\tilde{\mu} & = & \frac{\mu}{2} + \frac{1}{2}\sum_{j>0} 
                  v_{\rm DH}[a j] \nonumber \\
\tilde{v}_{\rm DH} & = & \frac{1}{4} v_{\rm DH}
\end{eqnarray}
and $a$ is the nearest-neighbor distance between dissociable groups 
on a straight line.

\subsection{Non-uniform mean-field approach with two sublattices}

\subsubsection{Mean-field equations}

In principle, the above statistical one-dimensional problem can be solved
using transfer-matrix techniques which take the long-ranged interactions into
account via a multiple-time integration with a suitably chosen kernel.
In order to obtain a simple, tractable
solution we use mean-field methods, which are implemented in the 
following way.
The Gibbs variational principle can be used to obtain
an upper bound for the free energy $F = -{\rm ln} Z$,
\begin{equation}
F \leq F_0 + \left<\mathcal{H}\right>_0 - \left<\mathcal{H}_0\right>_0
\label{Gibbs}
\end{equation}
In this inequality $\mathcal{H}_0$ is a trial Hamiltonian 
(to be specified below) and $F_0 = -{\ln}Z_0$, 
where $Z_0$ is the partition function obtained from $\mathcal{H}_0$; 
The thermal averages in Eq.~(\ref{Gibbs}) are evaluated using 
$\mathcal{H}_0$. We introduce
the trial Hamiltonian
\begin{equation}
\mathcal{H}_0 = h_0\sum_{i}\tilde{s}_{2i}+h_1\sum_i\tilde{s}_{2i+1}
\end{equation}
which separates the polymer into two sublattices. The variational
parameters $h_0,h_1$ are fields which act on the charges 
in the two sublattices. By minimizing  
the right hand side of Eq.~(\ref{Gibbs})
with respect to $h_0$ and $h_1$,
the following equations are obtained,
\begin{eqnarray}
h_0 & = & \tilde{\mu} + J \left<\tilde{s}_0\right>_0 
+ K\left<\tilde{s}_1\right>_0 
\nonumber \\
h_1 & = & \tilde{\mu} + J \left<\tilde{s}_1\right>_0 
+ K\left<\tilde{s}_0\right>_0 
\label{meanfield}
\end{eqnarray}
where
\begin{equation} 
\left <\tilde{s}_0\right>_0 = -{\rm tanh} (h_0)
\ \ \ ; \ \ \ 
\left<\tilde{s}_1\right>_0 = -{\rm tanh} (h_1), 
\label{tanh1}
\end{equation}
and
\begin{equation}
J = \frac{1}{2} \sum_{j>0}v_{\rm DH}[2ja] \ \ \ \  ; \ \ \ \ 
K = \frac{1}{2} \sum_{j\geq 0}v_{\rm DH}[(2j+1)a]
\label{JKdef}
\end{equation}

The choice of two sublattices (as opposed to more sublattices with
a larger period) is related to the strong anti-correlation that can
exist between adjacent monomers, and will be further motivated below.

\subsubsection{Main properties of mean-field equations}
Equations (\ref{meanfield}) and (\ref{tanh1}) 
always have a symmetric solution for which 
$h_0 = h_1$. However,
the symmetric solution is not always the minimum of the free energy but
can be, instead, a saddle point. In 
these cases two other solutions exist, both of which
break the symmetry between the two sublattices, \textit{i.e.},
$h_0 \neq h_1$. One solution can be obtained from the other by 
exchanging $h_0$ and $h_1$. The average charging degree of the polymer 
is then equal to:
\begin{equation}
\left< s\right>_0 = \frac{\left<\tilde{s}\right>_0+1}{2} =
\frac{\left<\tilde{s_0}\right>_0 + \left<\tilde{s}_1\right>_0 + 2}{4} 
\end{equation}

In order to
understand for which parameters symmetry breaking occurs, 
let us consider first the case $\tilde{\mu} = 0$. 
In this case the Hamiltonian exhibits the  symmetry
$\tilde{s}_i \rightarrow -\tilde{s_i}$ in addition
to the symmetry of exchanging the two sublattices. Even
if the latter symmetry is broken, we have
$\left<\tilde{s}\right>_0 = 0$, or equivalently
$\left<s\right>_0 = 1/2$, \textit{i.e.}, 
exactly half of the monomers are dissociated. Using the fact
that $h_0=-h_1$, Eqs.~(\ref{meanfield}) and (\ref{tanh1})
reduce in this case to one transcendental equation,
\begin{equation}
h_0 = (K-J)\,{\rm tanh} (h_0)
\end{equation}
This equation has a non-zero solution
(where $h_0 \neq h_1$) only if:
\begin{equation}
K-J > 1
\label{condition}
\end{equation}
If this condition is met, a sublattice symmetry breaking solution 
also exists within a certain range
of $\tilde{\mu}$ values around zero. Outside the range
$\tilde{\mu} = \pm \tilde{\mu}_c$ there is no symmetry 
breaking, \textit{i.e.}, 
$\left<\tilde{s}_0\right>_0 = \left<\tilde{s}_1\right>_0$.
If condition (\ref{condition}) 
is not met, there is no symmetry breaking solution
for any value of $\tilde{\mu}$.

The solution with $h_0 = h_1$ (no sublattice symmetry breaking) can be found
by substituting this equality in Eqs.(\ref{meanfield}) and (\ref{tanh1}), 
leading to the transcendental
equation 
\begin{equation}
h_0 = (K+J)\,{\rm tanh} (h_0)
\end{equation}
In a uniform mean-field approximation this solution is found 
for all values of $\tilde{\mu}$, whereas in our case it 
applies only for $|\tilde{\mu}| \ge \tilde{\mu}_c$. 

Before considering concrete examples we comment 
on the nature of the transition at
$\tilde{\mu} = \pm \tilde{\mu}_c$. This 
transition is second order, \textit{i.e.},  
$\left<\tilde{s}_1\right>_0 - \left<\tilde{s}_0\right>_0
\rightarrow 0$ as $\tilde{\mu} \rightarrow \pm \tilde{\mu}_c$.
Note that a non-zero $\tilde{\mu}$ does not break the symmetry of
exchanging the two sublattices in the Hamiltonian. This is
why the order parameter is continuous at the transition.
However the derivative of the order parameter
with respect to $\tilde{\mu}$ is discontinuous and diverges when 
approaching the transition from the side where symmetry breaking
occurs. 
Similarly, the derivative of the average dissociation degree 
with respect to $\tilde{\mu}$ has a discontinuity at the
transition. These are
artifacts of the mean-field approach, since the exact solution 
for a one dimensional system with short-ranged interactions
cannot exhibit a real phase 
transition \cite{Huang}. 
However, our non-uniform mean-field scheme still predicts the 
average charge very accurately, 
as will be demonstrated below. 
\subsection{Uniform dielectric constant}
In the case of a uniform dielectric constant, in which
the screened interaction is given by Eq.~(\ref{vDH}), 
the summations in Eq.~({\ref{JKdef}}) can
be performed explicitly, yielding

\begin{eqnarray}
J & = & -\frac{l_B}{4a}\left[{\rm ln}\left(1-{\rm e}^{-\kappa a}\right)
+{\rm ln}\left(1+{\rm e}^{-\kappa a}\right)\right] \nonumber \\
K & = & -\frac{l_B}{4a}\left[{\rm ln}\left(1-{\rm e}^{-\kappa a}\right)
-{\rm ln}\left(1+{\rm e}^{-\kappa a}\right)\right] \\
\tilde{\mu} & = & \frac{\mu}{2} 
- \frac{l_B}{2a}\,{\rm ln}\left(1-{\rm e}^{-\kappa a}\right)
\end{eqnarray}
The condition for sublattice symmetry breaking, Eq. (\ref{condition}),
 translates to
\begin{equation}
K-J = \frac{l_B}{2a}\,{\rm ln}\left(1+{\rm e}^{-\kappa a}\right) > 1
\end{equation}
Increasing $\kappa$ decreases $K-J$ and thus the possibility for sublattice
symmetry breaking. 
Symmetry breaking can occur for a finite range of $\kappa$ and pH only if the
condition
\begin{equation} \label{condition2}
\frac{l_B}{a} > \frac{2}{\rm ln 2} \simeq 2.9
\end{equation}
is met.
For $l_B = 7.0$\,\AA~this
condition leads to $a \lesssim 2.4$\,\AA. Hence vinyl-based polymers 
with an acid group on every second Carbon atom such as poly-styrene-sulfonate
or poly-acrylic-acid with a charge-distance of 
$a \approx 2.5$ \AA\ are marginally close to symmetry breaking within the present model. 
However, a dielectric discontinuity
due to the polymer backbone can increase $K-J$ considerably,
as will be discussed in the following sub-section. 

Examples with symmetry breaking will be shown in the following
sub-section, while here we restrict ourselves to the case
of dielectric continuity and no symmetry breaking.
In the case of no symmetry breaking all the dependence 
on $\kappa$ enters through the quantity
\begin{equation}
K+J = - (l_B/2a)\,{\rm ln}\left(1-{\rm e}^{-\kappa a}\right),
\end{equation}
which increases with increasing  Debye length $\kappa^{-1}$.

\subsubsection{Results}
\begin{figure*}
\centerline{
\scalebox{0.45}{\includegraphics{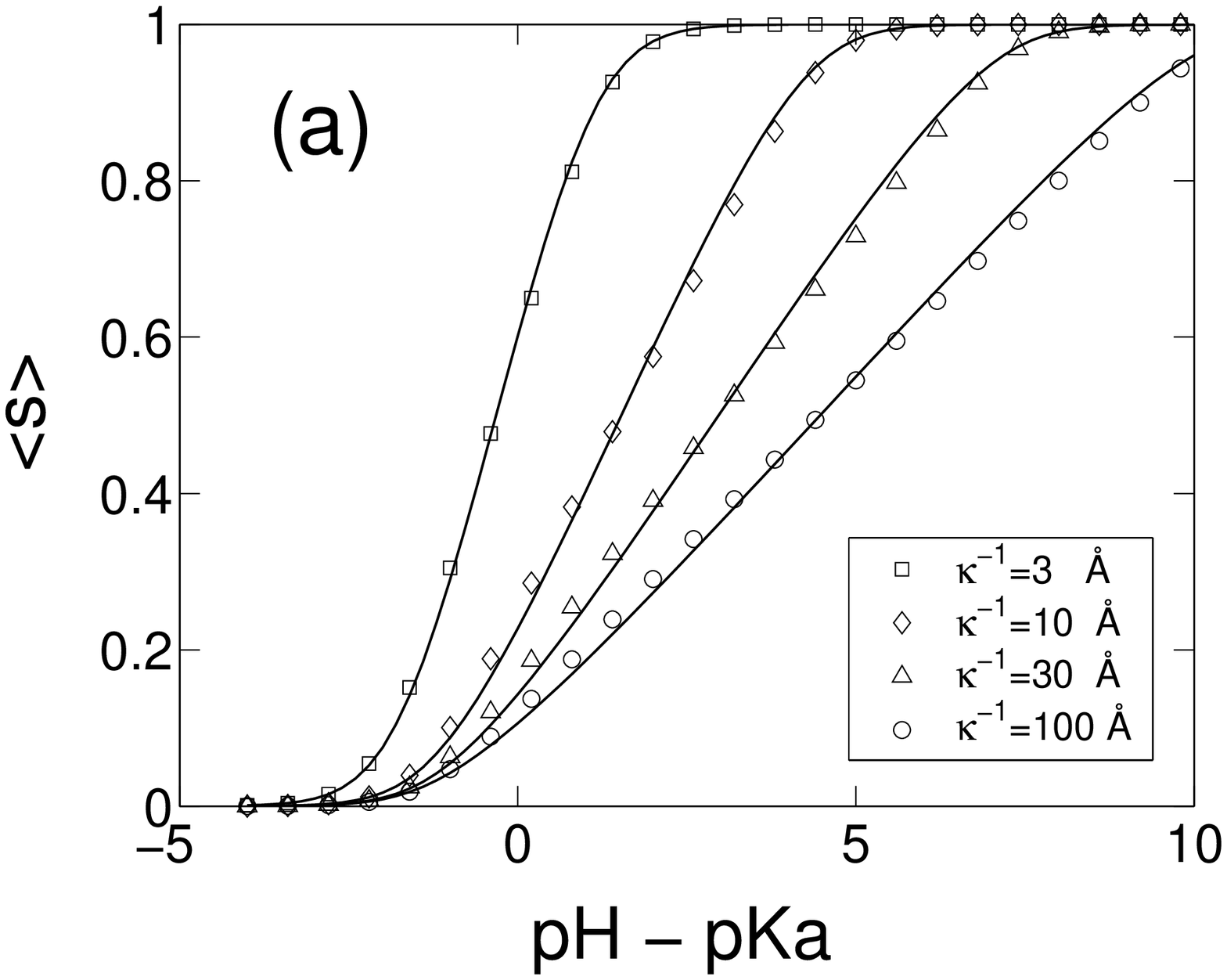}}
\hspace{0.2cm}
\scalebox{0.45}{\includegraphics{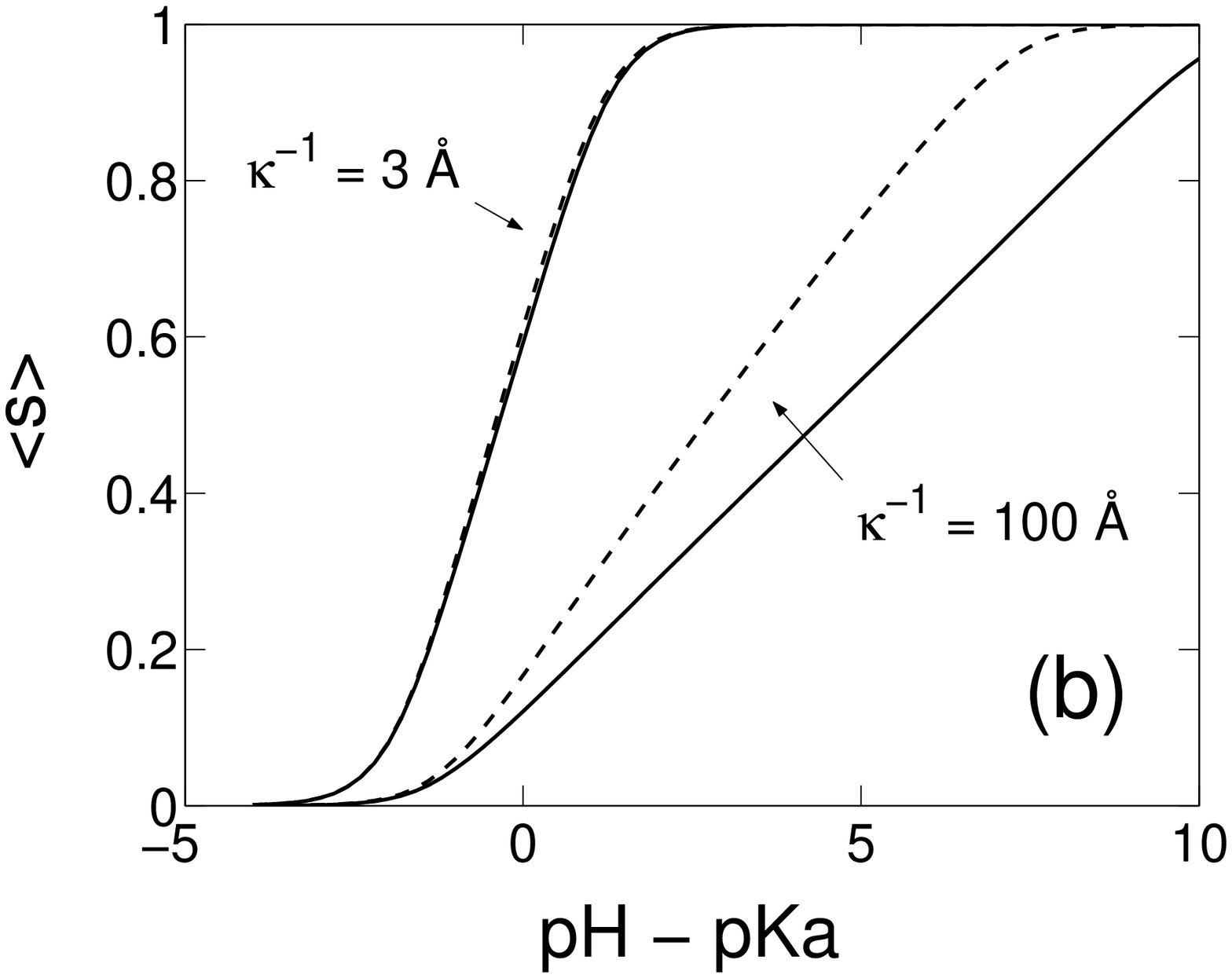}}
}
\caption{(a) Average degree of dissociation 
of a polyacid as function of pH-pKa, calculated using a 
mean-field approximation (solid lines). For comparison an exact enumeration
over all configurations is also shown (symbols) with $N=20$ 
and using periodic boundary conditions. 
Results for four different values of the Debye length are shown,
$\kappa^{-1} = $ 3\,\AA~(squares), 10\,\AA~(diamonds), 30\,\AA~(triangles)
and 100\,\AA~(circles). The separation between charged
groups is $a = 2.5$\,\AA. (b) Average degree of dissociation as
function of pH-pKa, calculated using an exact enumeration. The solid
lines show enumeration results with periodic boundary conditions 
(as shown using symbols in part (a)). These results are compared
with enumeration without periodic boundary
conditions (dashed lines). Two different values of the Debye length
are shown in the plot, $\kappa^{-1} = $ 3 (to the left) and 100\,\AA
(to the right).
}
\end{figure*}

In Fig.~1(a) we show the average degree of charge dissociation 
following from our mean-field equation for
a polyacid with $a = 2.5\,$\,\AA, for four different 
salt concentrations, corresponding to $\kappa^{-1}$ = 3, 10,
30 and 100\,\AA\ (solid lines). 
These results are compared with an exact 
calculation of the free energy for a finite chain with $N=20$ dissociable groups
(symbols),
by enumeration over all $2^N$ states. 
For the exact enumeration,
periodic boundary 
conditions are imposed by setting the interaction between
monomers $i$ and $j$ to be
\begin{equation}
v_{\rm DH}^p(i,j) = \sum_{n = -\infty}^{\infty} v_{\rm DH}(i, j + nN).
\end{equation}
The comparison between mean-field and the exact enumeration
is very good for all four values of $\kappa^{-1}$ shown in 
the figure. Note that in all these cases there is no symmetry breaking
in the mean-field solution, as expected since the charge distance
of  $a=2.5$\,\AA\ does not satisfy the condition Eq. (\ref{condition2}). 
Comparison of the four curves shows that $\kappa^{-1}$
has a large effect on the degree of charging. As $\kappa^{-1}$ is
increased each monomer interacts more strongly with the other
monomers. This increased repulsion reduces the charging, or, 
as one might put it, the long-ranged repulsion between charged
groups makes even strong PEs weak.

Throughout this work we will assume that the polymer is long compared
to the Debye length. However, it is important to realize that for
shorter polymers the average degree of dissociation depends on
the polymer length. In order to demonstrate this point we show
in Fig.~1(b) the average degree of dissociation as function of pH for
a finite chain of length $Na$, where $N=20$ and $a = 2.5$\,\AA. 
In the calculation an exact enumeration over all 
configurations is performed, without periodic boundary conditions 
(dashed lines). 
The results are compared with an enumeration with periodic boundary 
conditions (solid lines), as was done in Fig.~1(a). For 
$\kappa^{-1} = 3$\,\AA, the polymer length
is larger than the screening length, $Na \gg \kappa^{-1}$ and the two 
calculations yield nearly identical results. In the second
case shown in the plot, $\kappa^{-1}=100$\,\AA,
$\kappa^{-1}$ and $Na$ are of the same order of magnitude
and there are significant finite size effects. These results may
be important for the interaction of short DNA oligomers with substrates,
as they show that the polymer length affects adsorption behavior 
also via the effective charge of PEs.

\subsubsection{Restriction to nearest-neighbor interactions}

A common approximation that was previously applied for 
the charge regulation of PEs is to consider only nearest-neighbor (NN) 
interactions \cite{Marcus,HarrisRice,Lifson}. 
The reason is that exact closed-form solutions are available in this case.
Within the non-uniform mean-field approach, this approximation corresponds to 
setting $K = (l_B/2a)\,{\rm exp}(-\kappa a)$ and $J = 0$. 
Note that the combination $K+J$, which determines the 
solution without symmetry breaking, 
 is smaller in the NN case than in full interaction case. 
On the other hand the combination $K-J$, which 
affects the symmetry breaking (see Eq.~(\ref{condition})), 
is larger in the NN case. 

\begin{figure}
\centerline{
\scalebox{0.45}{\includegraphics{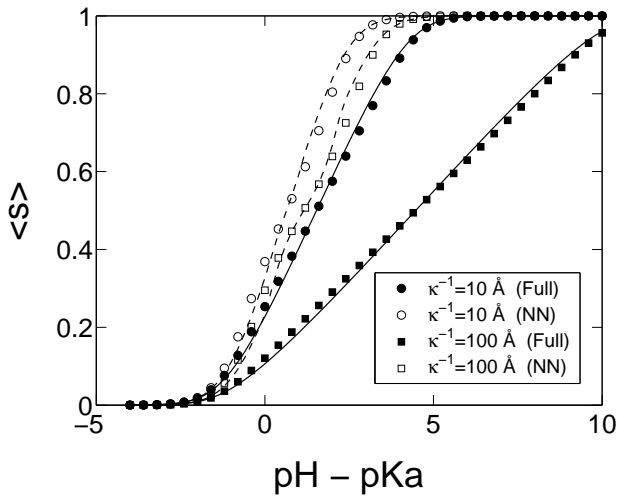}}
}
\caption{Comparison of a nearest neighbor 
mean-field approximation (dashed lines) with a mean-field
calculation taking into account
all long range interactions (solid lines) for a polyacid having
$a = 2.5$\,\AA, as function of pH-pKa. Results for 
two values of the Debye
length are shown. They can be distinguished in the
plot according to the type of symbols (circles, $\kappa^{-1}=10$\,\AA;
squares, $\kappa^{-1} =$ 100\,\AA). The symbols show
exact enumeration results with $N=20$ and periodic boundary 
conditions (empty symbols - only nearest neighbor interactions; 
full symbols - full long range interaction). 
}
\end{figure}
In Fig.~2 we compare the NN predictions 
(dashed lines and open symbols) with those obtained using the full
long-range interaction (solid lines and filled symbols),
for two different values of the Debye length. The symbols
show exact enumeration results, obtained using periodic boundary conditions 
and are thus representative of an infinitely long system,
while the lines show mean-field results. For both values
of the Debye length there are significant deviations
between the NN result and the full interaction.
As can be expected, these deviations are larger for the 
larger screening length, $\kappa^{-1} = 100$\,\AA, since in this
case the interaction between further-nearest neighbors contributes
significantly to the total interaction.
Note that for $\kappa^{-1}=100$\,\AA \
and NN interactions (open square symbols and
dashed line) there is also a small effect of symmetry breaking
in the mean-field solution, resulting in a non-monotonous
slope near $\left<s\right>=0.5$. A similar effect is seen
in the exact enumeration. As a main result of this section,
we see that the restriction to nearest-neighbor interactions is in 
general not a good approximation, 
while the mean-field approach reproduces the exact enumeration 
results very accurately.

\subsection{Non-uniform dielectric constant}

\begin{figure}
\centerline{\scalebox{0.7}{\includegraphics{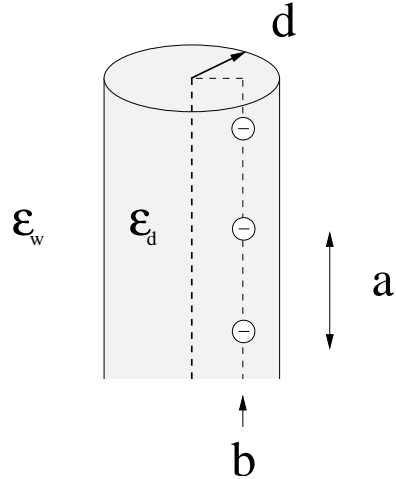}}}
\caption{Schematic representation of a simple
model taking into account the difference between the dielectric 
properties of water and a polymer's backbone.
A PE is modeled as a cylinder of radius $d$
with a dielectric constant $\varepsilon_d$, while
the dielectric constant outside the cylinder 
is equal to $\varepsilon_w$. Charged
groups that can dissociate from the polymer are located at 
regular intervals $a$ from each other, at a distance $b$ from
the cylinder axis. 
}
\end{figure}
So far we neglected effects of the dielectric discontinuity
between the polymer and its surroundings. 
As these effects tend to increase the electrostatic interactions,
they are expected to be important for the dissociation process.
They
were estimated in Ref.~\cite{Borkovec} 
using a simple model, shown in
Fig.~3. The PE is modeled as a cylinder of
radius $d$ and dielectric constant $\varepsilon_d < \varepsilon_w$, 
where $\varepsilon_w \simeq 80$ is the dielectric constant of 
water. Charged groups are assumed to be equally spaced along
the cylinder axis, with separation $a$. Here we 
generalize this model to some extent by placing these charged 
groups at a distance $0 \leq b \leq d$ from the axis, as shown 
in Fig.~3. 
The electrostatic potential exerted by one such charge on
another one was calculated in Ref.~\cite{Borkovec}
and is given by
\begin{equation}
\psi = \frac{\varepsilon_w}{\varepsilon_d}
\frac{l_B}{z}
+ \frac{1}{2\pi}\sum_{n=-\infty}^{\infty}\,W_p(z,n)
\label{psicyl}
\end{equation}
where $z$ is the distance between the charges. 
The first term is equal to the
electrostatic interaction within a medium of dielectric constant
$\varepsilon_d$, with no screening by salt. In the second term,
$W_p$ is equal to
\begin{equation}
W_p(z,n) = 4\frac{\varepsilon_w}{\varepsilon_d}l_B
\int_{0}^{\infty}{\rm d}k\,{\rm cos}(k z)
\left[I_n(k b)\right]^2 R(k,n)
\end{equation}
where
\begin{widetext}
\begin{equation}
R(k,n) = 
\frac{k \varepsilon_d\left[K_{n-1}(kd)+K_{n+1}(kd)\right]K_n(pd)
-p\varepsilon_w\left[K_{n-1}(pd)+K_{n+1}(pd)\right]K_n(kd)}
{k\varepsilon_d\left[I_{n-1}(kd)+I_{n+1}(kd)\right]K_n(pd)
+p\varepsilon_w\left[K_{n-1}(pd)+K_{n+1}(pd)\right]I_n(kd)}
\end{equation}
\end{widetext}
$p = \left(k^2+\kappa^2\right)^{1/2}$ and $K_n, I_n$ are the $n$-th
modified Bessel functions of the first and second kind, respectively.

In Ref.~\cite{Borkovec} only the case $b = 0$ was 
considered. For two charges located exactly at the polymer
axis, $b = 0$, the electrostatic interaction approaches 
$(\varepsilon_w/\varepsilon_d)l_B/z$ at short separations, 
while for larger separations it crosses
over to the interaction in the aqueous ionic solution, 
$l_B{\rm exp}(-\kappa z)/z$. 
The former interaction is typically much larger than the latter, 
leading to a two-step charging curve
and failure of a uniform mean-field approach.
On the other hand, only close-by monomers interact strongly with
each other, motivating the use of a nearest-neighbor (NN) model
\cite{Borkovec},
where only interactions between neighboring monomers 
are taken into account. The free energy and average degree of
dissociation 
can then be calculated exactly \cite{Marcus,HarrisRice,Lifson}, 
\textit{e.g.}, using the transfer 
matrix method \cite{Baxter,Huang}.

The NN approximation indeed predicts the two-step behavior of the 
charging curve, but it can fail for large values of the Debye 
screening length, since in this case further-nearest neighbor
interactions become important. 
For large $\kappa^{-1}$ long-range interactions can be 
important although they are much weaker than the interactions
between neighboring monomers, as shown in the following
numerical examples.

\subsubsection{Results for radially symmetric charge distribution}

\begin{figure*}
\scalebox{0.45}{\includegraphics{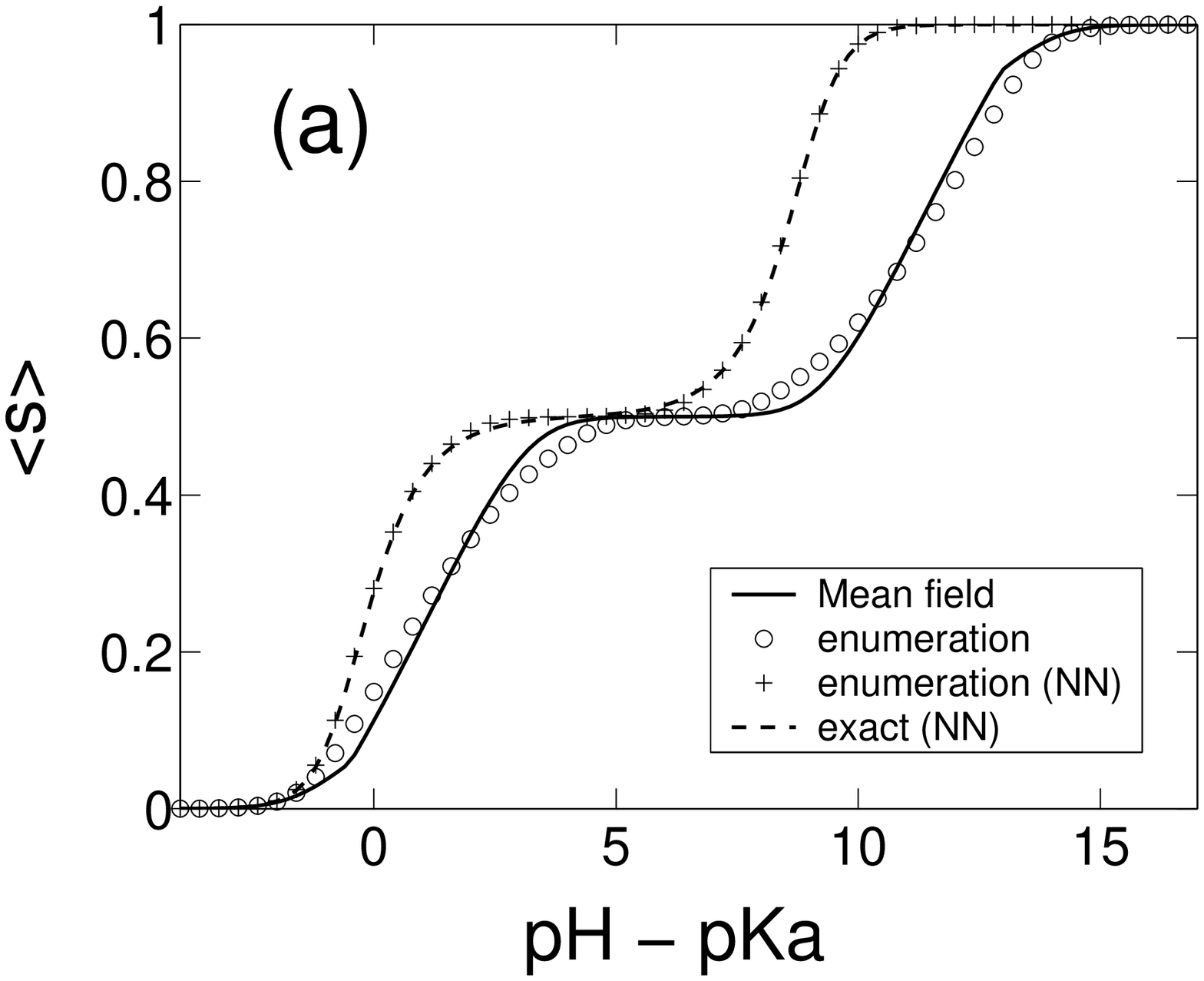}}
\hspace{0.2cm}
\scalebox{0.45}{\includegraphics{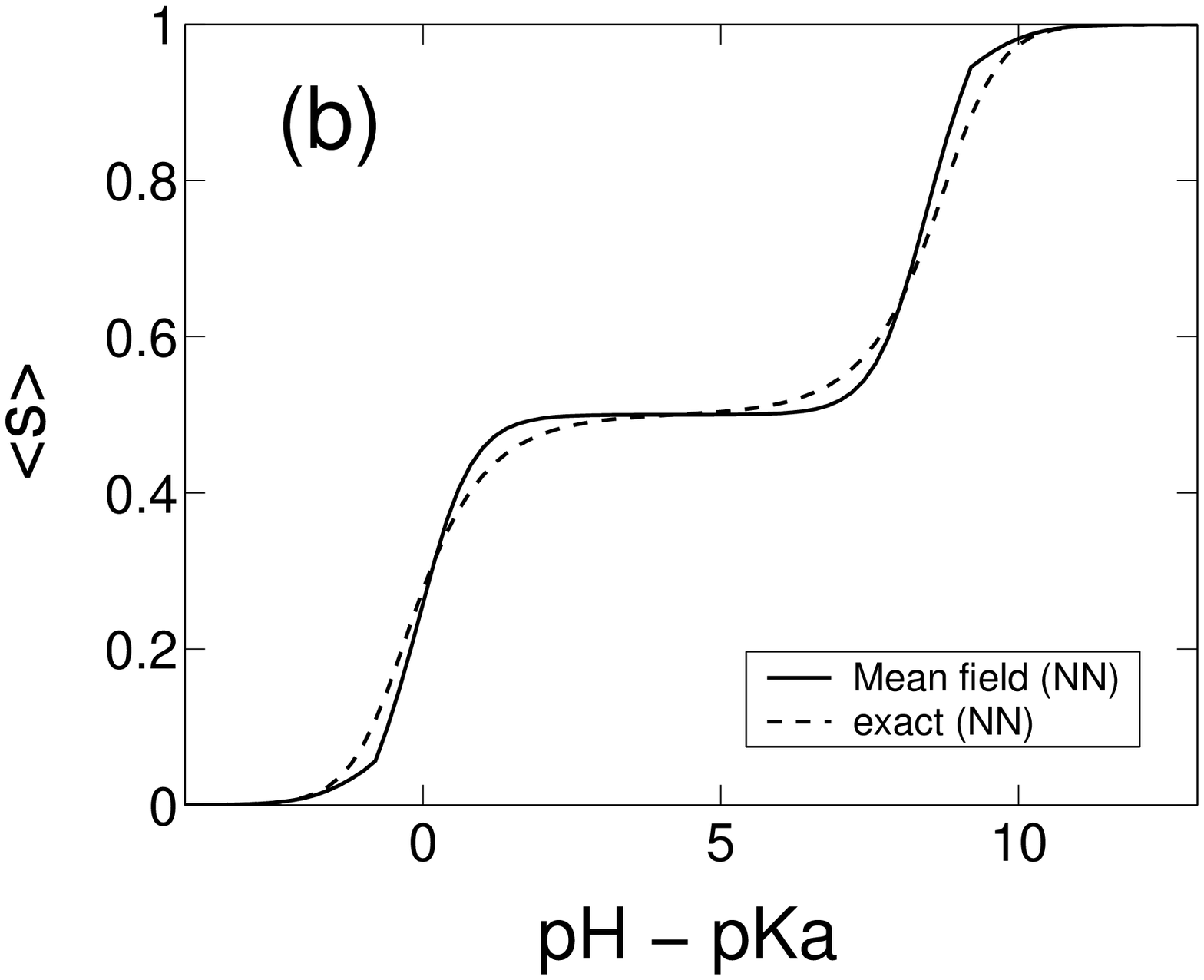}}
\\
\vspace{0.2cm}
\scalebox{0.45}{\includegraphics{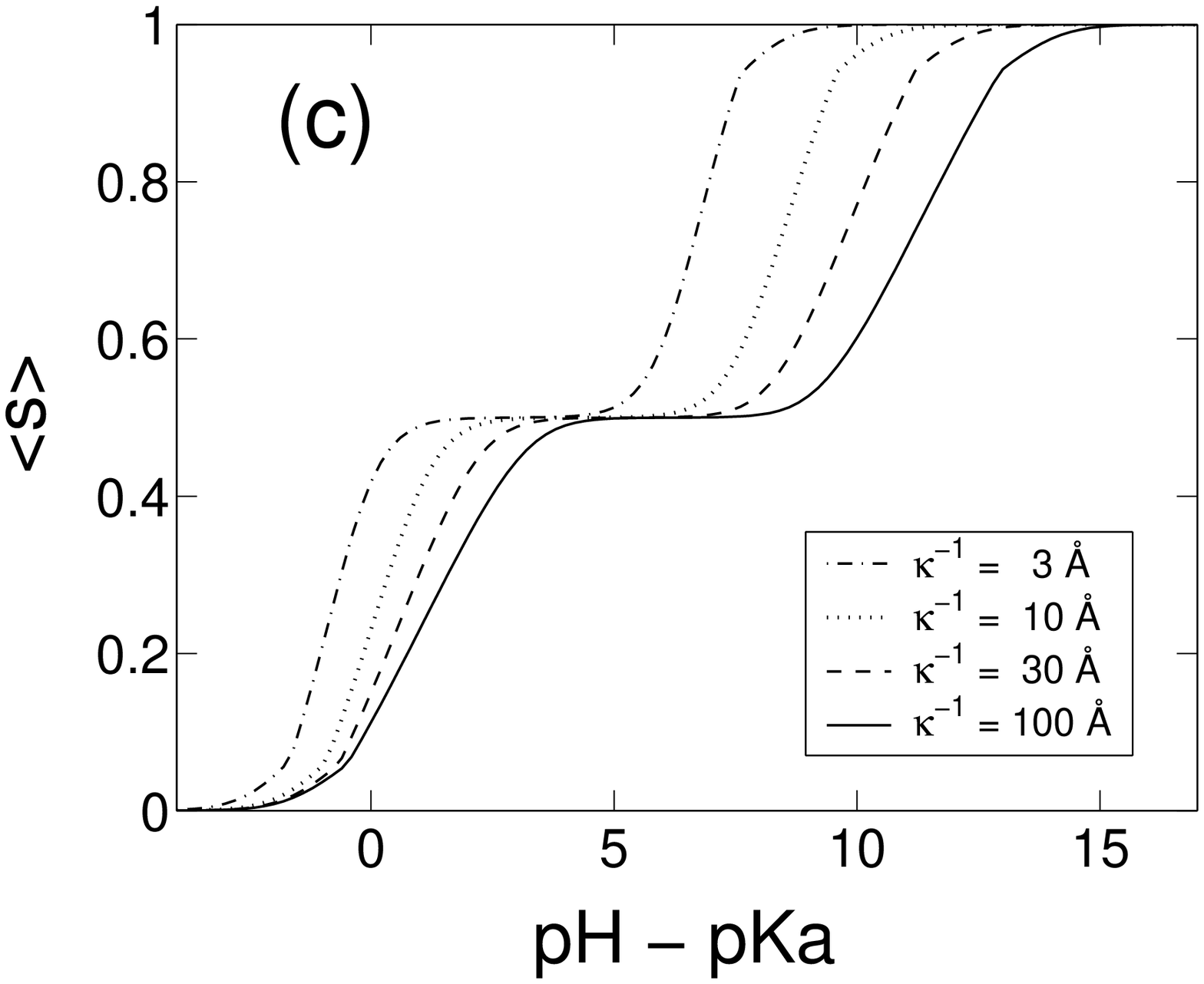}}
\caption{Average degree of dissociation of
a polymer with strong interactions between close by monomers,
characterized by a plateau near $\left<s\right>=1/2$. The 
interactions between monomers are calculated using the cylindrical
model shown in Fig.~3, with $\varepsilon_d = 3$, 
$\varepsilon_w = 80$, $d = 2.5$\,\AA, $a = 3.5$\,\AA,
$b = 0$ and a Debye length $\kappa^{-1}=100$\,\AA. 
(a) Mean-field results with two sublattices (solid line) 
are compared with
an enumeration over all configurations with $N=20$ and periodic
boundary conditions (circle symbols). 
The mean-field results show two cusps, where transitions occur between
a solution with no symmetry breaking and one with symmetry breaking
(the latter occurs for intermediate pH). 
Results are also compared with 
an enumeration taking only nearest neighbor (NN) interactions
into account (crosses) and the exact solution with NN interactions,
calculated using the transfer matrix method (dashed line).
(b) Mean-field calculation with two sublattices taking only
NN neighbor interactions into account (solid line), compared with
the exact solution with NN interactions (dashed line).  
(c) Dissociation as function of PH for four different values of
the Debye screening length: 3, 10, 30 and 100\,\AA, 
calculated using the mean-field approximation with two sublattices. 
All parameters other than $\kappa$ are as in part (a). The interaction
parameters $J,K$ used in the mean-field approximation are 
equal to $0.21,4.36$ ($\kappa^{-1}$=3\,\AA); $0.46,4.86$ 
($\kappa^{-1}$=10\,\AA); $0.87,5.30$ ($\kappa^{-1}$=30\,\AA);
and $1.37,5.84$ ($\kappa^{-1}$=100\,\AA).
}
\end{figure*}
In Fig.~4(a) the NN prediction (dashed line, exact solution)
is compared with an enumeration using the full long range 
interaction (circles).
The Debye length is $\kappa^{-1} = 100$\,\AA~and interactions between
monomers are calculated using Eq.~(\ref{psicyl}) with $b = 0$,
$d = 2.5$\,\AA, $\varepsilon_w/\varepsilon_d = 80/3$
and a monomer separation $a = 3.5$\,\AA~. 
In all calculations with dielectric discontinuity 
we use a monomer separation $a = 3.5$\,\AA~rather than $2.5$\,\AA,
which corresponds to a somewhat smaller fraction of dissociable groups.

The exact solution of the NN model in Fig.~4(a) (broken line)
deviates significantly from the enumeration results with the full range
of interactions included (circles). In contrast,
our non-uniform  mean-field approach with symmetry breaking
(solid line in Fig.~4(a)) is semi-quantitatively correct. 
The success of our generalized mean-field approximation 
is one of the main results in the first part of this work.
Note that the main difference with respect to enumeration is that
the mean-field approximation over-estimates the
effects of symmetry breaking, as seen from the exaggerated size of
the plateau region.

We also present in Fig.~4(a) a comparison between the NN exact solution
(dashed line) and an enumeration taking only nearest-neighbor 
interactions into account (crosses). 
This is done in order to test finite size effects 
in the exact enumeration. The enumeration and exact solution
yield almost identical results, demonstrating that
an enumeration with $N=20$ is typically very accurate for a single polymer.
We note that periodic boundary conditions 
are essential in order to obtain this 
level of accuracy in enumeration, when long range
interactions are included (compare Fig. 1b).
As another test for the enumeration procedure we 
increased $N$ to 30 for several different choices of 
$v_{\rm DH}(z)$. 
In all these cases deviations from results with $N=20$ 
were insignificant.

Figure 4(b) shows a comparison between the NN exact solution (broken line)
and a mean-field calculation with two sublattices (solid line), 
taking only the NN interaction into account,
for $\kappa^{-1} = 100$\,\AA~.
The NN-mean-field approximation shows an
effect similar to the curve in  Fig~4(a): 
two artificial discontinuities in the 
derivative of $\left<s\right>$ are seen, corresponding to two erroneous 
second-order transitions. In addition, the plateau is more pronounced 
compared to the exact solution. 
Nevertheless, the overall prediction for 
$\left<s\right>$ as function of pH is quite accurate.

Finally, the dependence on the Debye screening length $\kappa^{-1}$
is investigated in Fig.~4(c), using our non-uniform
mean-field approach. The parameters $J$ and $K$ of the sublattice
interactions are calculated for each value of $\kappa$ using 
Eqs.~(\ref{JKdef}) and (\ref{psicyl}). For all the four values
of $\kappa^{-1}$ that are shown, $\kappa^{-1}=3$, 10, 30 and
100\,\AA\ a pronounced plateau is visible. The cusps that are present
in all four cases are artifacts due to the mean-field approach.

\subsubsection{Dependence on the position of charged groups}

\begin{figure}
\centerline{\scalebox{0.45}{\includegraphics{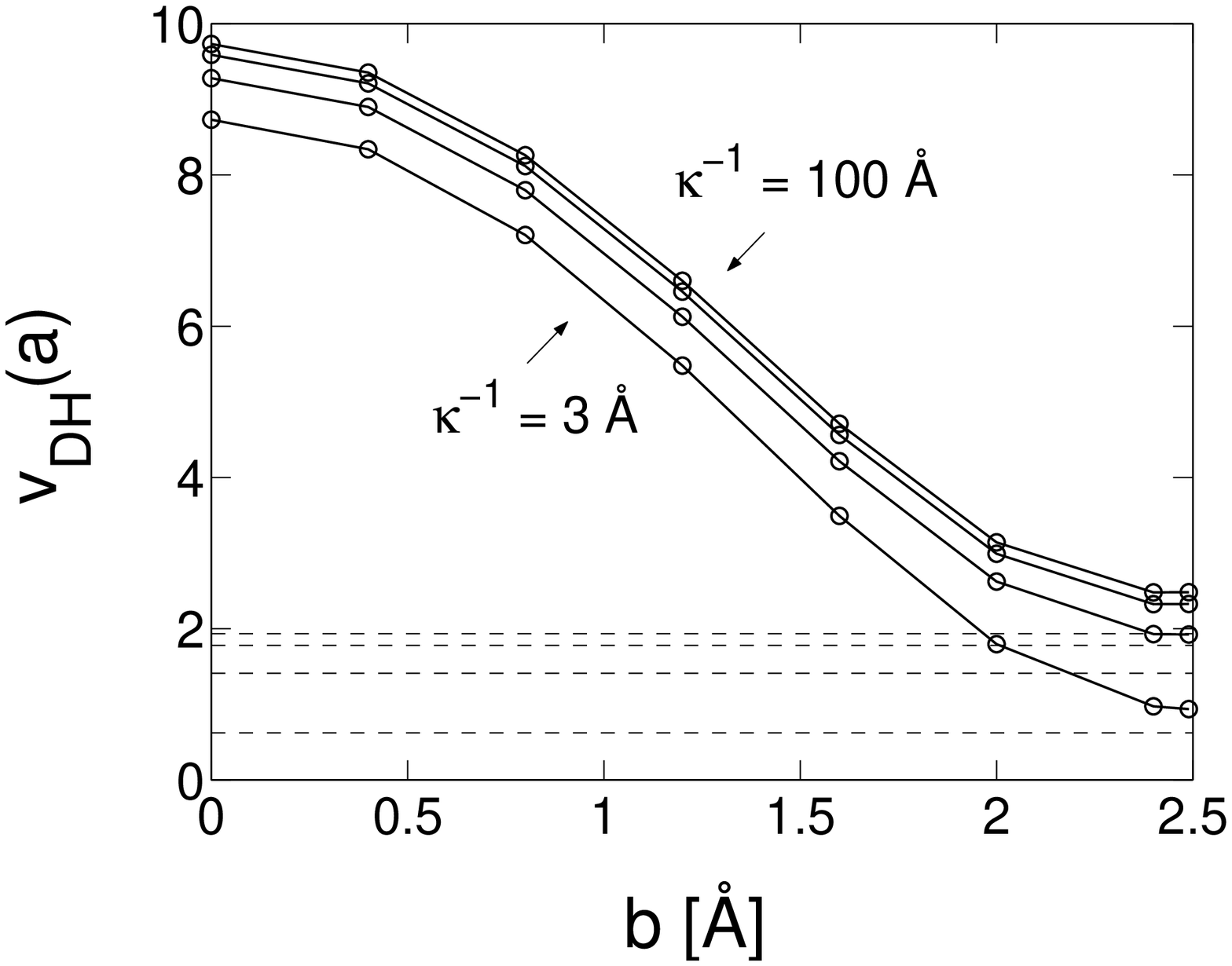}}}
\caption{Dependence of the electrostatic interaction
between two monomers on their distance $b$ from the polymer axis,
within the cylindrical model of Fig.~3. The distance between
the monomers is $a = 3.5$\,\AA~and the other model parameters are 
$d = 2.5$\,\AA, 
$\varepsilon_d$ = 3 and $\varepsilon_w$ = 80. Results are shown
for four values of the Debye length,
$\kappa^{-1} = $3, 10, 30, and 100\,\AA. The electrostatic
interaction in an aqueous ionic solution is shown 
for comparison using dashed lines.
}
\end{figure}
In most polyacids and polybases the charged units are located
in side groups, rather than being close to the polymer axis. 
This raises the question whether an
interaction much larger than the usual Debye-H\"uckel interaction
will occur even if the charges are displaced from the axis. 
Within the simple cylindrical model
presented above, this question can be addressed by calculating
the electrostatic interaction between two monomers 
as a function of $b$. Such a calculation is shown
if Fig.~5, for four different values of the screening length
$\kappa^{-1}$. The monomers are separated by a 
distance of $a = 3.5$\,\AA, while the polymer radius is taken
as $d = 2.5$\,\AA, as in Fig~4. In all four cases a very large
decrease of $v_{\rm DH}$ occurs with increase of $b$ toward
the cylinder boundary, $b = r$. For large $r$ this result
is not surprising, since the cylinder becomes similar to a planar
interface, separating an aqueous ionic solution and a low dielectric
medium. Near such an interface 
the electrostatic interaction is equal to twice the screened
electrostatic interaction in water \cite{Netz99}. For large cylinder
radius $d \gg \kappa^{-1}$ we have checked that Eq.~(\ref{psicyl}) 
indeed yields this result. In Fig.~5 the cylinder radius is not
large compared to the Debye length and the
interaction close to the cylinder boundary is even smaller
than in the planar limit. The screened
interaction in water is shown for comparison using dashed lines.

The above analysis demonstrates that actual electrostatic 
interactions between near-by monomers depend strongly on 
the spatial organization of the PE. These interactions
probably cannot be estimated reliably using simplified
models such as the cylindrical one presented above. 
The detailed polymer structure, as well as other effects such as the 
discreteness of the solvent, must be taken into account.

\subsection{Further discussion of the two-sublattice approximation}

Although the comparison with exact enumeration demonstrates that
a two-sublattice model is useful,
one may ask to what extent the separation into 
two sublattices has a physical
significance. In order to discuss this question we note that 
within the plateau region of the titration curve
there is typically a strong anti-correlation between
even and odd monomers.
In order to understand this anti-correlation
one may think of
the ground state of the Hamiltonian in Eq.~(\ref{znew}), concentrating 
first on the special case $\tilde{\mu} = 0$.
When the interaction between monomers favors 
opposite dissociation values, the ground state is
typically a periodic array of alternating values in the even and
odd positions, $s = +1$
and $s = -1$. The long
range order of the ground state is not preserved within the
exact theory at any finite
temperature, due to the entropy associated with domain boundaries
in a one dimensional system \cite{Huang}. Nevertheless, at a certain
range of pH values around 
$\tilde{\mu} = 0$ we may expect a staggered correlation 
function with strong anti-correlation 
between even and odd monomers.

\begin{figure}
\centerline{\scalebox{0.45}{\includegraphics{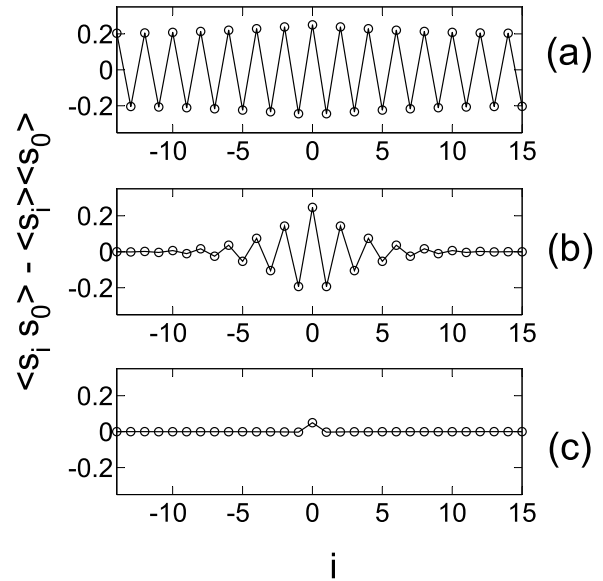}}}
\caption{Correlation function between the
dissociation of a monomer and that of its neighbors,
$\left< s_i s_0\right> - \left<s_i\right>\left<s_0\right>$, 
calculated from an
exact enumeration over all configurations of a PE having 
$N=30$ monomers, with periodic boundary conditions. 
All physical parameters are as in Fig.~4(a). The pH is
equal to 7 in (a), corresponding to $\tilde{\mu} = 0$,
to 9 in (b), and to 13.5 in (c).
}
\end{figure}
As an example, Fig.~6 shows the correlation function, 
$\left< s_i s_0\right> - \left<s_i\right>\left<s_0\right>$, 
for a PE having the same parameters as in
Fig.~4(a), calculated by 
exact enumeration over all states of a PE of length $N=30$
and using periodic boundary conditions. Results are shown
for three different pH values: in the top plot
pH = 7, corresponding to $\tilde{\mu} = 0$.
The correlation function has a staggered form which persists
over the full length of the PE. Nevertheless it is clear that there
is no true long range order because the (anti) correlation
decreases slightly with monomer separation. In the middle 
plot, where 
the pH is equal to 9, the correlation has a shorter range,
persisting only up to a distance
of about 8 sites from the center monomer. Note that
a pH value of 9 is approximately at the right edge
of the plateau region seen in
Fig.~4(a). With further increase of pH the
correlation length continues to decrease and
at pH = 13.5 (bottom plot) there is almost no correlation
even between adjacent monomers. This pH
value is close to the transition point that is found in
the two-sublattice
model, beyond which there is no symmetry breaking between the 
two sublattices. In summary, Fig.~6 demonstrates that the 
two-sublattice
approximation captures an essential physical property of the
dissociation pattern that is absent in the uniform mean field
approach, namely a strong anti-correlation between even and 
odd sites.

In principle, a periodicity other than two may be included in the 
formulation of the mean field equations, and could lead,
for certain parameters,
to a lower free energy than the two-fold periodicity. 
In such cases a plateau would be  expected in
the titration curve at an average degree of dissociation other than one
half. Comparison with the exact enumeration and with typical 
experimental
results indicates that such additional symmetry breaking into
structures with more than two sublattices does not occur within
the physical parameters considered in this work.

In the second part of this work, where we will look at
the interaction between two weak polyelectrolytes, we will employ
simple Debye-H\"uckel interactions, as well as
the interaction within a cylindrical dielectric cavity
with $b=0$, which constitute the two extreme cases. 
In both cases we expect 
a mean-field approach with two sublattices to be
adequate in order to predict the average charging, 
as was demonstrated in the preceding discussion.

\section{Interaction between polyacid and polybase}

\subsection{Uniform mean-field approach}

\begin{figure}
\centerline{\scalebox{0.70}{\includegraphics{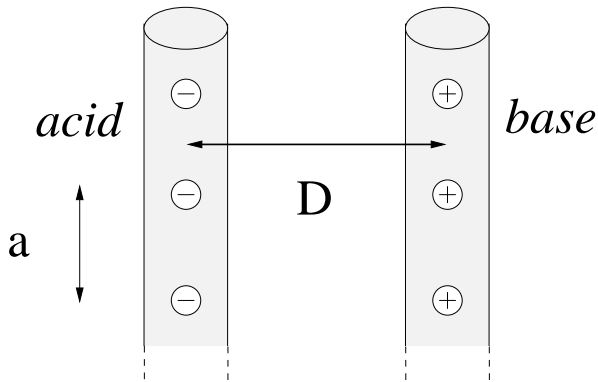}}}
\caption{Schematic illustration of a model
describing interaction of a polyacid with a parallel polybase,
separated by a distance $D$. The distance between charged
groups in both of the PEs is $a$. 
For simplicity the charged groups are facing each other.
}
\end{figure}
The model we consider is shown schematically in Fig.~7. 
A polyacid (left) and polybase (right) 
are aligned parallel to each other and separated by a distance $D$. 
For simplicity we assume that the distance between charged groups
(denoted by $a$) is identical in the two polymers and that
the charge lattice are in phase with  each other in the two polymers, as
shown in the figure. We would like to calculate the average charge
on the two polymers and the free energy as a function of $D$. 

We consider first the case of a uniform  dielectric 
constant, Eq.~(\ref{vDH}), and also assume
that for each polymer a uniform mean-field theory
(with no symmetry breaking) is adequate. As was shown in the calculations
for a single polymer,
the latter assumption is justified for monovalent
monomers having a nearest neighbor separation 
$a \gtrsim 2.5$\,{\AA}.

It is convenient to define for the polyacid
\begin{equation}
s_a = \frac{1+\tilde{s}_a}{2}
\label{sadef}
\end{equation}
and for the polybase
\begin{equation}
s_b = \frac{1-\tilde{s}_b}{2}
\label{sbdef}
\end{equation}
where $s_a$ and $s_b$ are zero for an uncharged monomer and one
for a charged (dissociated) one.
With these definitions both $\tilde{s}_a$ and $\tilde{s}_b$ 
increase with pH. 
The mean-field equations are found in a similar way as in the
single polymer case, and are given by
\begin{eqnarray}
h_a & = & \tilde{\mu}_a
            + J \left<\tilde{s}_a\right>_0
            + K \left<\tilde{s}_b\right>_0
\nonumber \\    
h_b & = & \tilde{\mu}_b 
            + J \left<\tilde{s}_b\right>_0
            + K \left<\tilde{s}_a\right>_0
\label{meanfield2}
\end{eqnarray}
where 
\begin{equation}
\left<\tilde{s}_a\right>_0 = -{\rm tanh}(h_a) 
\ \ \ ; \ \ \  
\left<\tilde{s}_b\right>_0 = -{\rm tanh}(h_b). 
\label{tanh2}
\end{equation}
The coefficients in these equations are given by:
\begin{eqnarray}
J & = & \frac{1}{4}\sum_{i\neq 0}v_{\rm DH}(ia) \nonumber \\
K & = & \frac{1}{4}\sum_i v_{\rm DH}\left[
\sqrt{(ia)^2+D^2}\right] \nonumber \\
\tilde{\mu}_a & = & -\frac{2.303}{2}({\rm pH}-{\rm pKa})+\Delta\tilde{\mu}
\nonumber \\
\tilde{\mu}_b & = & -\frac{2.303}{2}({\rm pH}-{\rm pKb})-\Delta\tilde{\mu}
\label{JKtwo}
\end{eqnarray}
where
\begin{equation}
\Delta\tilde{\mu} = \frac{1}{4}\sum_{i\neq 0}v_{\rm DH}(ia)
-\frac{1}{4}\sum_i v_{\rm DH}\left(\sqrt{(ia)^2 + D^2}\right)
-\frac{l_B\kappa}{2}
\label{defs2}
\end{equation}
Equations~(\ref{meanfield2})-(\ref{tanh2}) are very similar to 
Eqs.~(\ref{meanfield})-(\ref{tanh1}), with a number
of important differences. First, the subscripts $a$ and $b$
do not represent two sublattices but instead 
distinguish between the polyacid and polybase. Another difference
is that the chemical potentials
$\tilde{\mu}_a$ and $\tilde{\mu}_b$ are usually not equal to each
other. Most importantly, $J$ is almost always larger
than $K$, whereas for the two sublattice case 
$J$ is smaller than $K$. It is easy to show that
for $J > K$ Eqs.~(\ref{meanfield2})-(\ref{tanh2}) 
have a single solution.

\subsubsection{Results}

\begin{figure}
\centerline{\scalebox{0.45}{\includegraphics{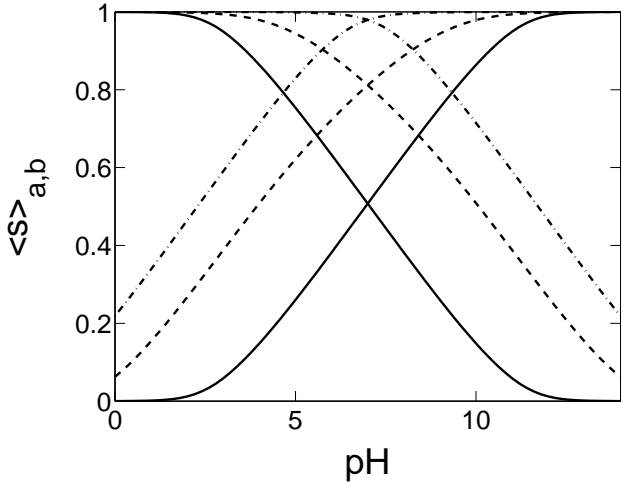}}}
\caption{Average dissociation degree of 
a polyacid (increasing lines) interacting with a 
polybase (decreasing lines), as function of pH.
The distance between charges is $a = 2.5$\,\AA; 
pKa = 4, pKb = 10 and the Debye length
is $\kappa^{-1} = 30\,$\,\AA. Results are shown for three values
of the inter-polymer separation: $D = 100$\,\AA~(solid lines),
$10$\,\AA~(dashed lines) and $5$\,\AA~(dash-dot lines). The symmetric
case where ${\rm pH} - {\rm pKa} = {\rm pKb} - {\rm pH}$ occurs at pH = 7.
}
\end{figure}
\begin{figure*}
\centerline{
\scalebox{0.45}{\includegraphics{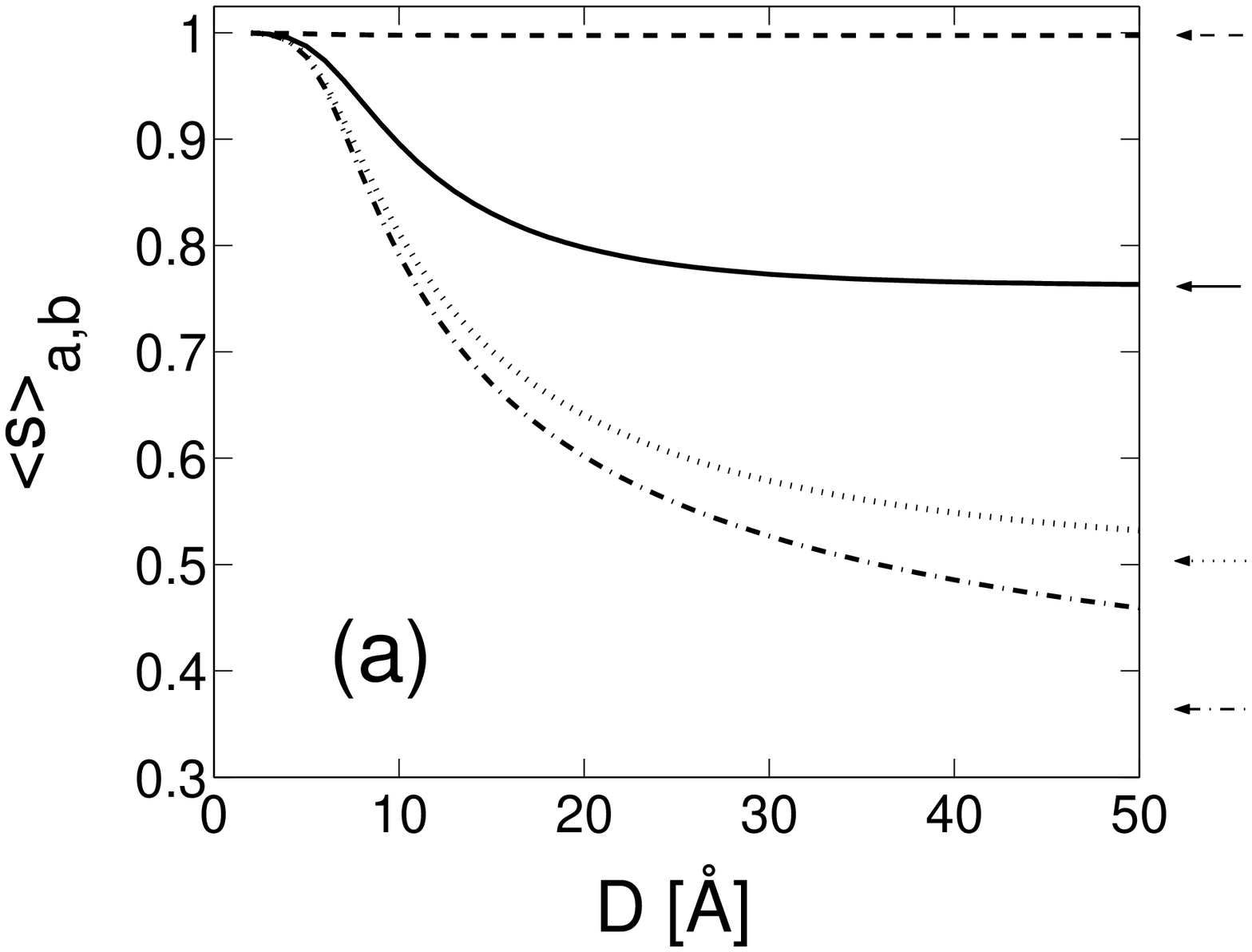}}
\hspace{0.2cm}
\scalebox{0.45}{\includegraphics{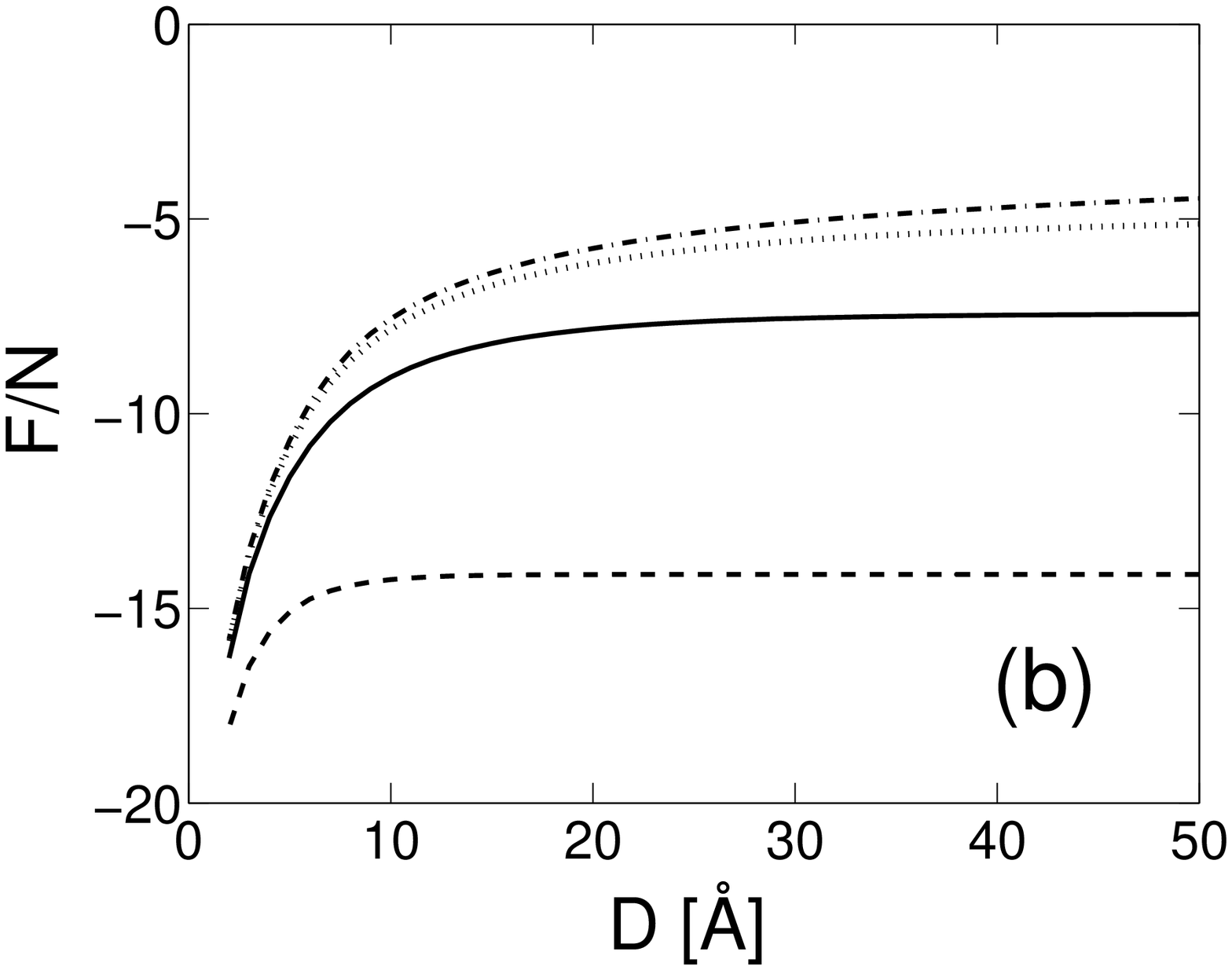}}
}
\caption{Average degree of dissociation (a)
and free energy per monomer, $F/N$ (b) as function of the distance
$D$ between a polyacid and polybase, with 
${\rm pH}-{\rm pKa} = {\rm pKb}-{\rm pH}=3$.
Simple Debye-H\"uckel interactions
are used with $\kappa^{-1} = $ 3\,\AA~(dashed line), 10\,\AA~(solid 
line), 30\,\AA~(dotted line) and 100\,\AA~(dash-dot line). 
All other parameters are as in Fig.~8. The arrows on the right hand
side of (a) show the value of $\left< s\right>$ for an isolated PE.
}
\end{figure*}
The electrostatic interaction
between the polyacid and polybase increases dissociation in both
polymers (in 
contrast to the interactions within each PE,
which inhibits charged groups from dissociating). Figure 8 shows
the degree of charging of a polyacid and polybase having 
$a = 2.5$\,\AA, as function of pH and for three different values
of $D$. When the polymers are sufficiently far away from each other 
their dissociation curves are identical to those of a single polymer.
For smaller separation the average charge increases.
An important case occurs when the pH is tuned such that
\begin{equation}
\left({\rm pH}-{\rm pKa}\right) = -\left({\rm pH}-{\rm pKb}\right)
\label{symmetric}
\end{equation}
For example, with the parameters used in Fig.~8, 
pKa = $4$ (similar to poly-acrylic-acid) and pKb = $10$ 
(similar to poly-vinyl-amin),  
this equality holds at ${\rm pH} = 7$. In this case, one has
$\tilde{\mu}_b = -\tilde{\mu}_a$, as seen from Eq.~(\ref{JKtwo}), 
and the solution of Eqs.~(\ref{meanfield2})-(\ref{tanh2}) has the properties
$h_a = -h_b$ and 
$\left<\tilde{s}_a\right>_0 = -\left<\tilde{s}_b\right>_0$.
Using the definitions in Eqs.~(\ref{sadef})-(\ref{sbdef}),
the average charging degrees of the polyacid
and polybase are then equal to each other, and the value of $h_a$ is found
from the single transcendental equation:
\begin{equation}
h_a = \tilde{\mu}_a + (K-J)\,{\rm tanh}(h_a)
\label{symmetriceq}
\end{equation}

In the following examples we restrict ourselves to the symmetric case 
described by
Eqs.~(\ref{symmetric}) and (\ref{symmetriceq}). 
Figure 9(a) shows the average degree of dissociation
as function of the polymer separation $D$ (identical for the polyacid
and polybase). Results are shown for 
$a = 2.5$\,\AA, ${\rm pH}-{\rm pKa} = {\rm pKb}-{\rm pH} = 3$ and for
four different values of the Debye length, ranging between $3$\,\AA~and
$100$\,\AA. When $D$ is large compared to $\kappa^{-1}$ the
polymers do not interact, and their average charge is equal to
its value in an isolated polymer (compare with Fig.~2 at pH-pKa = 3).
This value depends strongly on $\kappa^{-1}$.
At separations $D$ of order $\kappa^{-1}$ and smaller, the average charging
increases with decrease of $D$ and approaches unity (full dissociation) at 
contact.

We turn to the free energy of the two interacting PEs, shown 
in Fig.~9(b). In this figure the free energy $F$ is divided by $N$, 
the number of  monomers in each PE. A distinctive feature in this
figure is that $F$ is almost independent on $\kappa^{-1}$ at small
separations. In contrast to this
short separation behavior, $F$ depends strongly on $\kappa^{-1}$ at large
separations. In order to understand these two behaviors we consider each one
of the two limits separately:

\subsubsection{Small PE distances, $D \ll \kappa^{-1}$}

At short separations the average degree of dissociation
saturates and is independent on $\kappa^{-1}$,
as can be seen in Fig.~9(a). The electrostatic interaction
energy of the two polymers also becomes nearly independent
on $\kappa^{-1}$, as can be understood from the following
argument. Consider the two polymers as uniformly, oppositely 
charged and parallel lines. In the limit
of no screening, $\kappa = 0$, the electrostatic energy 
is dominated by interactions at distances of order $D$ and smaller. 
At distances larger than $D$ opposing positive and negative 
charges in the two polymers can be regarded as dipoles and 
the electrostatic interaction between them
decays as $1/z^3$ where $z$ is their distance, measured 
parallel to the polymers. 
As long as $\kappa^{-1} \gg D$ screening
affects only these dipole-dipole interactions, but not the main
electrostatic contribution coming from interactions at 
distances smaller than $D$. The independence of
both $\left<s\right>$ and the electrostatic energy 
on $\kappa^{-1}$ leads to the behavior seen at these 
small separations.

\subsubsection{Large PE distances, $D \gg \kappa^{-1}$}

\begin{figure}
\centerline{\scalebox{0.45}{\includegraphics{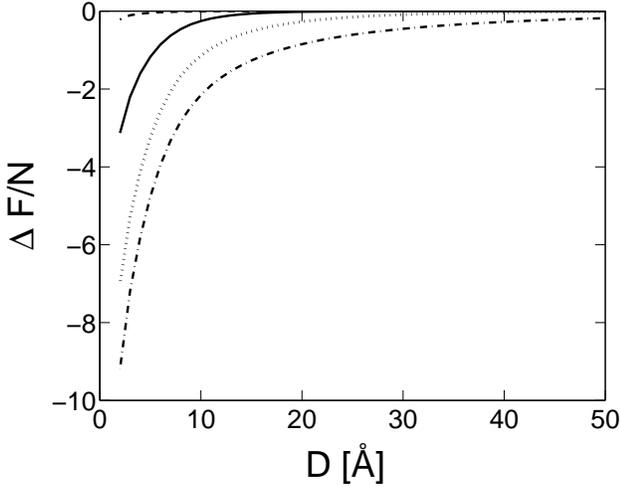}}}
\caption{The difference $\Delta F$ between
the free energy $F$ of a polyacid interacting with a polybase,
and the asymptotic form of Eq.~(\ref{approx0}). Four values
of $\kappa^{-1}$ are shown. These values, and all other parameters
and notations are as in Fig.~9. For each value of $\kappa^{-1}$ 
the values of $F_0$ and $s_0$ in Eq.~(\ref{approx0}) are equal
to the free energy and average degree of dissociation of
an isolated PE, respectively.
}
\end{figure}
At large separations the free energy approaches the sum
of free energies of the two isolated polymers. More precisely,
the average degree of dissociation on the two polymers approaches
a constant and the free energy can be approximated as follows:
\begin{equation}
\frac{F}{N} \approx \frac{2 F_0}{N}
- 2 l_B\left(\frac{s_0}{a}\right)^2 K_0(\kappa D)
\label{approx0}
\end{equation}
where $F_0$ and $s_0$ are the free energy 
and average degree of dissociation of a single, isolated PE,
respectively. These constants are unrelated
to the interaction between the two PEs but depend strongly
on $\kappa$. The modified Bessel function $K_0(\kappa D)$
characterizes the electrostatic interaction between  
two parallel and uniformly charged rods:
\begin{eqnarray}
K_0(\kappa D) & = & 
\int_{0}^{\infty}{\rm d}z\,
\frac{{\rm exp}
\left(-\kappa \sqrt{z^2+D^2}\right)}{\sqrt{z^2+D^2}}
\nonumber \\
& = & \int_{0}^{\infty}{\rm d}u\,
\frac{{\rm exp}\left(-\sqrt{u^2 + \kappa^2 D^2}\right)}
{\sqrt{u^2 + \kappa^2 D^2}} 
\label{U0}
\end{eqnarray} 
Deviations from the asymptotic form (\ref{approx0}) 
are expected to occur only 
when the average degree of dissociation deviates from $s_0$. 
This happens approximately when $D \lesssim \kappa^{-1}$
as can be seen in Fig.~10, where the difference between
$F/N$ and Eq.~(\ref{approx0}) is plotted for the same
four values of $\kappa^{-1}$ as in Fig.~9.

\subsection{Non-uniform mean-field approach}

In the case of stronger interactions between monomers within each polymer, 
the dissociation curve of a single 
PE is characterized by a plateau. In this case we expect a symmetry 
breaking transition with two sublattices on each PE. 
In order to deal with this case the mean-field Hamiltonian can be generalized
to account for sublattices on each one of the two PEs:
\begin{equation}
\mathcal{H}_0 = h_0^a\sum_{i}\tilde{s}^a_{2i}+h_1^a\sum_i\tilde{s}^a_{2i+1}
    + h_0^b\sum_{i}\tilde{s}^b_{2i}+h_1^b\sum_i\tilde{s}^b_{2i+1}
\end{equation}
The mean-field equations and free energy are found
using the Gibbs variational principle, in a similar fashion as
for the single polymer case. For example, the equation
for $h_0^a$ is:
\begin{equation}
h_0^a = \tilde{\mu}^a 
            + J \left<\tilde{s}_0\right>_0^a
            + K \left<\tilde{s}_1\right>_0^a
            + I_0 \left<\tilde{s}_0\right>_0^b
            + I_1 \left<\tilde{s}_1\right>_0^b
\label{example}
\end{equation}
where $\left<\tilde{s}_i\right>^\alpha_0 = -{\rm tanh}\,h_i^\alpha$.
Similar equations are obtained for $h_1^a$, $h_0^b$ and
$h_1^b$. The coefficients $J$,$K$,$I_0$ and $I_1$ in 
Eq.~(\ref{example}) are equal to:
\begin{eqnarray}
J & = & \frac{1}{4}\sum_{i\neq 0}v_{\rm DH}(2ia) 
\nonumber \\
K & = & \frac{1}{4}\sum_i v_{\rm DH}\left[2(i+1)a\right]
\nonumber \\
I_0 & = & \frac{1}{4}\sum_iv_{\rm DH}\left[
\sqrt{(2ia)^2 + D^2}\right]
\nonumber \\
I_1 & = & \frac{1}{4}\sum_iv_{\rm DH}\left[
\sqrt{(2i+1)^2 a^2+D^2}\right]
\label{JKI01}
\end{eqnarray}
and $\tilde{\mu}^a$, $\tilde{\mu}^b$ are given by 
Eqs.~(\ref{JKtwo}) and (\ref{defs2}). 
The four equations for $h_i^{\alpha}$
typically have multiple solutions; For example,
if symmetry breaking occurs on both polymers the number of solutions
is 9. 
In the limit of non-interacting polymers, $I_0=I_1=0$, 
four of these solutions are equivalent minima of the free energy,
related to each other by exchange of the two sublattices on
one or both of the polymers. Interactions between the polymers break 
the symmetry of exchanging only the sublattices in one of the polymers, 
and there are two (equivalent) global minima of the free energy. 

\subsubsection{Results}

As a concrete example we consider again the model shown in 
Fig.~3, which accounts for a low dielectric constant of the polymer
backbone. Parameters are similar to Fig.~4, $b = 0$, $a = 3.5$\,\AA, 
$d = 2.5$\,\AA~and $\kappa^{-1} = 100$\,\AA. The pH, pKa and pKb
values are chosen such that ${\rm pH}-{\rm pKa} = {\rm pKb}-{\rm pH} = 3$. 
The coefficients $J$ and $K$ are set as in Fig.~4 with 
$\kappa^{-1} = 100$\,\AA, $J = 1.4$ and $K = 5.8$. 
For the coefficients $I_0$ and $I_1$ we use
Eqs.~(\ref{JKI01}) with the screened Debye-H\"uckel interaction
in water, Eq.~(\ref{vDH}).
Both of these choices are approximations which become 
inaccurate when the polymers are very close to each other, since
the electrostatic Green's function should then be evaluated in
the presence of two dielectric cylinders. However we expect our results
to be qualitatively correct as will be further discussed below. 

\begin{figure}
\centerline{\scalebox{0.45}{\includegraphics{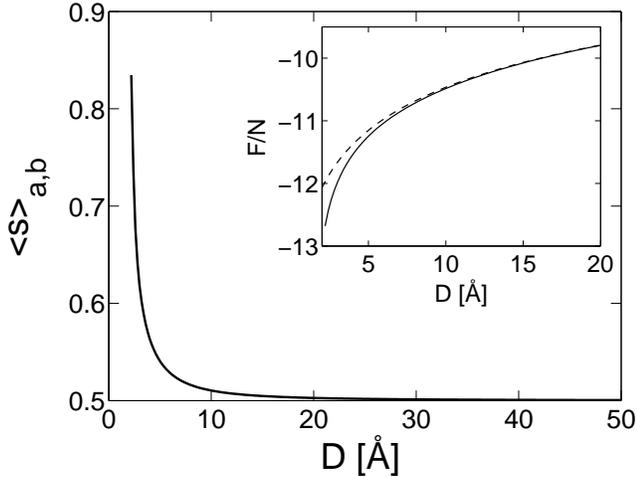}}}
\caption{Average degree of dissociation of
interacting polyacid and polybase, as function of their distance.
The interactions between monomers within each PE are
calculated assuming a low dielectric backbone, as in Fig.~3. The
parameters of the model are as in Fig.~4: $a = 3.5$\,\AA,
$d = 2.5$\,\AA, $b = 0$ and $\kappa^{-1} = 100$\,\AA, and
${\rm pH}-{\rm pKa}={\rm pKb}-{\rm pH}=3$. The inset shows the
free energy per monomer, $F/N$, as function of $D$ (solid line).
The dashed line shows the approximation of Eq.~(\ref{approx0})
with $s_0 = 1/2$ and $F_0$ set to match the value of $F$ at 
large $D$.
}
\end{figure}
Figure 11 shows the average charging of the two polymers as function
of their separation $D$. Due to the plateau in the dissociation curve
of each polymer, $\left<s\right>$ is close to $1/2$ in most of the
separation range. A sharp increase in $\left<s\right>$ is
found at close separations of a few Angstr\"oms. Note that this range,
where interactions between the polymers affect the average charging,
is much smaller than the Debye length,
$\kappa^{-1}=100$\,\AA. 

The inset shows the free energy (solid line)
as function of $D$. 
For comparison we show (by a broken line)
 the approximation of Eq.~(\ref{approx0}),
with $s_0$ equal to $1/2$ and $F_0$ matching the
value of $F$ at large $D$. 
We note that the constant $F_0$ used in Fig.~11 is not
exactly the free energy of an isolated PE. At very large separations,
when the PEs are truly isolated from each other, the average 
degree of dissociation is somewhat smaller than $0.5$ (see
Fig.~4 at pH-pKa = 3). At these separations the asymptotic form
of Eq.~(\ref{approx0}) applies with $F_0$ and $s_0$ set according
to the properties of the isolated PEs, while the form shown
in Fig.~11 matches the free energy at intermediate values of $D$ where
$\left<s\right> \approx 0.5$.
The free energy $F$ deviates from Eq.~(\ref{approx0}) 
only at very large separations, and at very small separations 
where $\left<s\right>$ is larger than $1/2$,
as seen in the inset of Fig.~11.
 
In summary, the symmetry-broken solution is stable for a wide range 
of distances and gives way to a symmetric solution only 
at very small distances.
With the dielectric discontinuity on both polymers taken properly
into account (affecting $I_0$ and $I_1$ as well as $K$ and $J$)
we can expect a stronger interaction between the polymers for small 
$D$. This will lead to an increase of the average charging
at somewhat larger values of $D$ than in Fig.~11.

\section{Summary}

The main result in the first part of this work concerns 
a generalization of the standard mean-field theory
of charge regulation in weak PEs.
The polymer is divided into two sublattices, allowing
explicitly for correlations between these sublattices
to be taken into account.  
Similar models have been studied in the past in the
context of Ising-like models. In
the Ising model, interactions are usually assumed to be
short-ranged. If only interactions between neighboring 
monomers are considered, the partition
function can be calculated exactly. For PEs the 
main advantage of using a mean-field approximation is that 
it allows long-range electrostatic interactions 
to be taken into account. 
Simultaneously, one expects mean-field methods to gain
in accuracy as the range of interactions increases.
 Our main result is that a 
mean-field approach with separation into two sublattices 
is adequate within  a wide range of model parameters. 
In particular it succeeds in the case of large 
inter-monomer interactions, where a uniform mean-field 
theory fails, while also taking into account 
long range interactions, which may still play an 
important role. 

A motivation for the use of a nearest-neighbor
approximation was recently suggested in Ref.~\cite{Borkovec}.
It was pointed out that a low dielectric
constant of the polymer backbone can lead to strong
enhancement of the coupling between close-by monomers.
We show that even within the model of Ref.~\cite{Borkovec},
the nearest neighbor approximation needs improvement
for large values of the Debye length, 
because of the contribution of
interactions between non-neighboring monomers.
On the other hand,
a mean-field approximation with two sublattices
is semi-quantitatively accurate. 
We also demonstrate that effects due to the  dielectric
discontinuity between the PE interior and the aqueous solvent
depend sensitively on the location of the charged groups within
the low-dielectric cavity; this is quite relevant, since
for most experimental PE architectures, 
the charged groups are not located centrally but are displaced towards
the aqueous interface.

The linearized Debye-H\"{u}ckel theory
 is used in this work to evaluate the interaction between 
monomers. This, of course, is only an approximation, whereas
in principle the full non-linear response
of the ionic solution must be taken into account.  
Use of Debye-H\"{u}ckel interactions is justified as long as the 
electrostatic potential is small compared to the thermal
energy. Hence this approximation is probably reasonably accurate
in the plateau region, 
where the average charge along the polymer is small.
Far away from the plateau, and for highly charged PEs, one needs
to go beyond Debye-H\"{u}ckel theory, using the non-linear 
Poisson-Boltzmann theory.
The great advantage of using pairwise interactions is that they
allow tractable, analytic solutions to be obtained. In contrast,
the nonlinear distribution of ions cannot be calculated analytically
even near a uniformly charged cylinder immersed
in a salt solution, let alone an inhomogeneously charged
PE. In light of this situation we believe that the results 
presented in this work provide a useful qualitative treatment
of charge regulation even for the case of highly 
charged PEs, although they may be quantitatively modified by 
charge renormalization due to nonlinear effects close
to the PE \cite{NetzOrland03}.

In the second part of this work we studied the interaction
between a polyacid and a polybase, using a mean-field
approximation. This interaction leads
to an increase of the average charging in both polymers
as they approach each other. 
In addition, the interaction energy between a weak polyacid and a weak
polybase is stronger than expected in the 
absence of distance-dependent charge regulation. 
This means that the effect of charge regulation
may be important for the building of stable multilayers,
since it allows to simultaneously decrease the repulsion 
between similarly charged weak PEs (since the charge 
regulation in this case decreases the charge strength) but
also leads to strongly bound polyacid-polybase pairs.
For close-by polymers the 
electrostatic energy is dominated by interactions 
between neighboring  monomers, and the free energy 
depends only weakly on the 
Debye screening length $\kappa^{-1}$. 
On the other hand at large separation 
between the polymers both the average degree of dissociation and
the free energy typically vary strongly with $\kappa^{-1}$.
The characteristic distance where interactions between the polymers
can affect their degree of dissociation is the Debye screening length. 
However, when there is a plateau in the charge \textit{vs.}
pH curve of a single PE, the degree of dissociation
may remain close to $1/2$ even at small separations compared to
$\kappa^{-1}$. 
In these cases a sharp increase in the average dissociation
can occur close to contact.

The increase of charging when weak polyacids and weak polybases 
come into contact could in principle be observed using infra-red 
spectroscopy in multilayers. However, one has to keep in mind
that in such highly concentrated systems the oppositely
charged groups will get very close to each other and form salt
bridges. 

\begin{acknowledgments}
YB wishes to thank the Minerva foundation for a grant.
This work was financially supported by Deutsche Forschungsgemeinschaft
(DFG, SFB 486) and the Fonds der Chemischen Industrie.

\end{acknowledgments}

\end{document}